\documentclass[prl,twocolumn,showpacs,preprintnumbers,amsmath,amssymb,footinbib]{revtex4-1}

\usepackage{graphicx}
\usepackage[colorlinks=true,citecolor=red,urlcolor=blue]{hyperref}
\usepackage{psfrag}
\usepackage{dcolumn}
\usepackage{bm}
\usepackage{color}
\usepackage{array}
\usepackage{longtable}
\usepackage{dcolumn}
 \def\braket#1{\mathinner{\langle{#1}\rangle}}

\def\mychi{\raisebox{0.35ex}{$\chi$}}

\input{Qcircuit}

\begin{document}

\date{\today}
\author{Simon E. Nigg and S.~M.~Girvin}
\affiliation{Department of Physics, Yale University, New Haven, CT
  06520, USA}
\begin{abstract}
We present a general protocol for stabilizer operator measurements
in a system of $N$ superconducting qubits. Using the dispersive
coupling between the qubits and the field of a resonator as well
as single qubit rotations, we show how to
encode the parity of an arbitrary subset of $M\leq N$ qubits, onto two
quasi-orthogonal coherent states of the resonator. Together with a fast
cavity readout, this enables the efficient
measurement of arbitrary stabilizer operators without locality constraints.
\end{abstract}
\title{Stabilizer quantum error correction toolbox for
  superconducting qubits}
\maketitle

Several milestones on the road to quantum
computing with superconducting circuits have recently been reached, such as the experimental
violation of Bell's inequality~\cite{Ansmann-2009a}
and the demonstration of rudimentary quantum error
correction (QEC)~\cite{Reed-2012a}. As the resources required
for more complete QEC protocols come within
experimental reach, it is desirable to develop a toolbox
sufficiently versatile to allow the implementation of a wide class of
codes. Most QEC codes can be described concisely in the stabilizer
formalism of Gottesman~\cite{Gottesman-Thesis}. In this framework a
QEC code is defined by
the subspace spanned by the eigenstates with eigenvalue $+1$ of a set
of commuting multi-qubit Pauli operators called stabilizer operators. Error
detection is achieved by measuring the stabilizer operators of the code; the
syndrome of an error being a sign flip of a subset of these operators. Correction in turn,
can be performed when the syndrome contains enough information to
identify the location and type of the error. The ability to measure
arbitrary multi-qubit Pauli operators would thus allow a direct
realization of stabilizer QEC codes, including non-local quantum low
density parity check codes~\cite{McKay-2004a,*Kovalev-2013a}.

Toric and surface codes~\cite{Kitaev-2003a,Bravyi-1998} defined on
two-dimensional qubit lattices are promising stabilizer codes with high
thresholds for fault-tolerance~\cite{Wang2-2010,*Fowler-2012a}. However
because the elementary (anyonic) excitations of these systems can
diffuse at no energy cost, quantum memories built from these codes
are thermally unstable~\cite{Dennis-2002a,Landon-Cardinal-2013a}. Thermal stability can be obtained by
engineering effective interactions between the
anyons~\cite{Chesi-2010a,Hutter-2012a} or by going to four dimensions, where deconfinement of anyons is energetically
suppressed~\cite{Alicki-2010a}. To be physically realizable however, the latter needs to be
mapped back onto a lattice of qubits with dimension $D\leq 3$. This
mapping inevitably leads to {\em non-local} stabilizer operators, which one
must be able to measure. In this work we take a first step in this
direction and propose a scheme to measure arbitrary stabilizer operators in a system of superconducting qubits off-resonantly coupled
to a common mode of a microwave resonator.

Several schemes for parity measurements of superconducting qubits
have recently been proposed~\cite{LaLumiere-2010a,DiVincenzo-2012a_notit,Tanamoto-2013a}. The
main advantage of our approach is the
ability to {\em selectively address an arbitrary subset of qubits}, without
the need for tunable couplings, in contrast to earlier work~\cite{Spiller-2006a,Myers-2007a}, and without restrictions on the
number of and distance between physical qubits defining a given
stabilizer operator. We thus
extend the superconducting qubit toolbox
with functionality similar to that
recently demonstrated for trapped
ions~\cite{Barreiro-2011a,*Muller-2011a}.

Central to our proposal is the off-resonant coupling between
a superconducting qubit and a single mode of a microwave
resonator~\cite{Blais-2004a} described by the dispersive Hamiltonian
$H_{\rm disp}=\mychi\,{\bm\sigma^z}{\bm a}^{\dagger}{\bm a}$, where ${\bm\sigma^z}=\ket{e}\bra{e}-\ket{g}\bra{g}$ is the Pauli
matrix in the computational basis $\{\ket{g},\ket{e}\}$ of the qubit and $\bm a$ (${\bm a}^{\dagger}$) denotes the photon annihilation
  (creation) operator of the cavity mode. This coupling describes a
  qubit-state-dependent frequency shift $\pm\mychi$ of the cavity, or equivalently a photon-number-dependent frequency
  shift $2n\mychi$ of the qubit. In
  the weakly dispersive regime $2\mychi\sim 1/T_2, \kappa$, where
  $\kappa$ is the bare cavity linewidth and $T_2^{-1}=(2T_1)^{-1}+\Gamma_{\phi}$ is
  the qubit coherence time composed of relaxation $1/T_1$ and pure
  dephasing $\Gamma_{\phi}$, this
  interaction enables a qubit-readout by measuring the phase-shift of transmitted or reflected
  microwaves~\cite{Blais-2004a}.
  In this work, we are interested in the ultra-strong
  dispersive regime of well-resolved resonances~\cite{Schuster-2007a},
  where $\kappa,T_2^{-1}\ll\mychi$. In this regime, we show
  how to encode the two
eigenvalues of an arbitrary multi-qubit Pauli operator onto quasi-classical oscillations of light
that differ in phase by $\pi$.
\begin{figure}[ht]
\includegraphics[width=0.9\columnwidth]{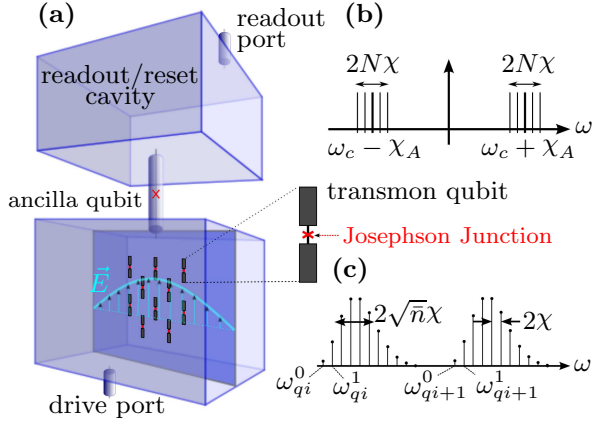}\caption{(Color
  online) {\bf (a)} 2D array of $N$ transmon qubits in a 3D
  cavity. The ancilla qubit and the upper cavity are used for readout/reset
  and manipulation purposes. Cavity {\bf (b)} and
  qubit {\bf (c)} spectra in the ultra-strong dispersive regime.\label{fig:1}} 
\end{figure}

Although our scheme is applicable to other types of superconducting
qubits, for clarity we
will frame our discussion around the specific case of transmon qubits. A transmon qubit~\cite{Koch-2007a,Paik-2011a} is formed by
a superconducting dipole-antenna with a Josephson
junction at its center with Josephson energy $E_J\gg E_C\equiv e^2/(2C_{\Sigma})$,
where $C_{\Sigma}$ represents the total capacitance between the antenna pads. Neglecting charge-dispersion
effects, which are suppressed exponentially in $E_J/E_C$~\cite{Koch-2007a},
the low-energy spectrum of an isolated transmon is well approximated by that of an
anharmonic oscillator with frequency $\omega_{01}\approx\sqrt{8E_JE_C}-E_C$
and weak anharmonicity $\omega_{01}-\omega_{12}\approx
E_C\ll\omega_{01}$. In state-of-the-art realizations, the qubit linewidth
$1/T_2$ is close to being limited by relaxation~\cite{Paik-2011a,
  Lenander-2011a, Catelani-2011a, Catelani-2012b}.
In this work we are interested in a
setup such as depicted in Fig.~\ref{fig:1} (a), where $N$ transmons are
coupled dispersively with strength $\mychi$ to a microwave field inside a 3D
cavity. For simplicity, we here discuss the case of equal dispersive couplings. In the
supplemental material we show how to cope
with the more realistic case of unequal dispersive shifts. For control and readout purposes, an ancilla
qubit, is further dispersively coupled to both the high-$Q$
cavity containing the $N$ qubits with $\mychi_A\gg N\mychi$ and to a low-$Q$
(readout) cavity, similar to the setup used in~\cite{Kirchmair-2013a}. We assume that both the ancilla qubit and the
readout cavity remain in the ground state,
except during readout and manipulation. Thus omitting, for now, the corresponding degrees of
freedom, we model this system in an appropriately rotating frame (see supplemental
material for details), by the effective Hamiltonian 
\begin{equation}\label{eq:3}
  H_0=\mychi\sum_{i=1}^{N}{\bm \sigma}_i^z{\bm a}^{\dagger}{\bm a}-K{\bm
  a}^{\dagger}{\bm a}^{\dagger}{\bm a}{\bm a}\,.
\end{equation}
The transmons are treated here as two-level systems assuming their
anharmonicity is larger than their linewidth (i.e.\
$E_C>1/T_2$). Furthermore, assuming the qubits to be
sufficiently detuned from each other, we neglect the cavity-mediated
qubit-qubit interaction. The latter leads to frequency shifts of
the order of $\mychi^2/\Delta$, where $\Delta$ is the detuning between
the two qubits. For the parameters used below ($\mychi=5\,{\rm MHz}$
and $\Delta\geq 2\,{\rm GHz}$), these shifts are
smaller than about $10\,{\rm KHz}$. The second term on the rhs of Eq.~(\ref{eq:3}) accounts for the
(negative) qubit-induced anharmonicity of the
cavity~\cite{Nigg-2012a,Bourassa-2012b}. In the weak dispersive
regime, this
term can usually be neglected as $K\ll\mychi$. We find that in the
ultra-strong dispersive regime it is necessary to account for its
leading order effect. We next show how to encode the parity ${\bm
  Z}_{S_N}=\bigotimes_{i=1}^N{\bm\sigma}_i^z$ of an $N$-qubit state
$\ket{\psi}_N$ onto two quasi-orthogonal
coherent states of the cavity differing in phase by $\pi$.

{\em Parity encoding}. Suppose the system is initially prepared in the product state
$\ket{\Psi}_{t=0}=\ket{\alpha}\ket{\psi}_N$, where $\ket{\alpha}$
is a coherent state
of the cavity with amplitude $\alpha$. Making use of the identity
$\exp[-i(\pi/2)\sum_{i=1}^N{\bm\sigma}_i^z]= (-i)^N{\bm Z}_{S_N}$, one
can show that under
the action of (\ref{eq:3}), at
time $T=\pi/(2\mychi)$, the state becomes
\begin{equation}\label{eq:4}
\ket{\Psi}_T={\bm U}_K\left(\ket{\alpha_N^{\ }}{\bm
  P}_{{S_N}}^++\ket{-\alpha_N^{\ }}{\bm P}_{{S_N}}^-\right)\ket{\psi}_N,
\end{equation}
where ${\bm P}_{S_N}^{\pm}=(\openone\pm {\bm Z}_{S_N})/2$ are the projectors
onto the even ($+$) and odd ($-$) qubit parity subspaces as measured
by the $\pm 1$ eigenvalues of the multi-qubit Pauli operator ${\bm
  Z}_{S_N^{\ }}$ and $\alpha_N^{\ }=(-i)^N\alpha$. Note that the
self-Kerr term is qubit-independent~\footnote{Qubit-state-dependent corrections to the self-Kerr are of order
  ${\hat\varphi}^6$ in an expansion of the Josephson
  potential $E_J\cos(\hat\varphi)$ and are smaller by a factor $\sim\sqrt{E_C/E_J}$.} and conserves the photon number ${\bm
  a}^{\dagger}{\bm a}$. Because it commutes with the
dispersive term, its effect factors out and is captured in
(\ref{eq:4}) by the unitary operator ${\bm U}_K=\exp[i\pi K/(2\mychi){\bm a}^{\dagger}{\bm
  a}^{\dagger}{\bm a}{\bm a}]$. 
For weak nonlinearity such that $\pi
K\ll \mychi$, the leading order effect of ${\bm U}_K$ acting on the coherent
states $\ket{\pm\alpha_N^{\ }}$ is a rotation of the mean amplitude by an
angle $\Delta\phi_{\rm nl}=\pi \bar n K/\mychi$ with the mean photon
number $\bar n=|\alpha|^2$. To leading order in
$K/\mychi$, the
state~(\ref{eq:4}) is thus well approximated by
\begin{equation}\label{eq:2}
\ket{\Psi}_T=\ket{\tilde\alpha_N^{\ }}{\bm
  P}_{{S_N}}^+\ket{\psi}_N+\ket{-\tilde\alpha_N^{\ }}{\bm P}_{{S_N}}^-\ket{\psi}_N\,,
\end{equation}
with $\tilde\alpha_N = \alpha_N^{\ } e^{-i\Delta\phi_{\rm nl}}$. The sub-leading order effect
is a damping of the mean amplitude by a factor
$\exp(-\Delta\phi_{\rm nl}^2/(2\bar n))$~\cite{Walls_Milburn-2008}. We emphasize that in
the ultra-strong dispersive regime $\kappa/\mychi\ll 1$ considered here, photon decay only
weakly damps the amplitude of the coherent states in~(\ref{eq:2}) by a
factor $\exp(-\kappa T/2)\approx 1-(\pi/4)(\kappa/\mychi)$. Ignoring these small
effects, we thus see that the dispersive interaction can be used to
encode the parity of the multi-qubit state onto two coherent states of the
cavity differing in phase by $\pi$.

{\em Subset selectivity}. Typically stabilizer operators are defined on subsets of qubits. Selectivity to $M\leq N$ qubits, labeled by the set $S_M^{\ }\subseteq S_N=\{1,\dots,N\}$,
can be achieved as follows. Consider the
identity
\begin{equation}\label{eq:5}
{\bm U}_{S_M^{\ }}(t)=\Big(\bigotimes_{i\notin S_M}{\bm \sigma}_i^x\Big){\bm
  U}_{S_N}\Big(\frac{t}{2}\Big)\Big(\bigotimes_{i\notin S_M}{\bm \sigma}_i^x\Big){\bm U}_{S_N}\Big(\frac{t}{2}\Big),
\end{equation}
where ${\bm U}_{S}(t)=\exp(-i t\mychi  {\bm a}^{\dagger}{\bm a}\sum_{i\in S}{\bm
  \sigma}_i^z)$.
Eq.~(\ref{eq:5}) can be easily shown using
${\bm\sigma}^x{\bm\sigma}^z{\bm\sigma}^x=-{\bm\sigma}^z$. Thus, by splitting the dispersive evolution of all $N$ qubits into two equal
halves and interspersing them with bit-flip operations on the qubits not
in $S_M$, we can effectively echo away the contribution of the latter to the total
``magnetization'' and implement the dispersive evolution ${\bm
  U}_{S_M^{\ }}(t)$ of the
qubits in $S_M^{\ }$ alone. Acting on an initial state of the form
$\ket{\alpha}\ket{\psi}_N$, ${\bm U}_{S_M}(T=\pi/(2\mychi))$ then encodes the
subset-parity ${\bm Z}_{S_M^{\ }}=\bigotimes_{i\in S_M^{\ }}{\bm \sigma_i^z}$ onto the state
of the cavity as explained above. The case of unequal
dispersive shifts is treated in the supplemental material.

Physically, the initial unconditional cavity displacement and bit-flips can be
implemented via fast microwave pulses (see
supplemental material). Because
of the dispersive
interaction, the qubit transition frequency of the $i$-th qubit splits into a ladder of
frequencies $\omega_{qi}^n=\omega_{qi}^0+2n\mychi$, corresponding to different photon
numbers in the cavity (Fig.~\ref{fig:1} (c)). The latter are Poisson distributed and peaked
around $\bar n$. Hence, to best approximate an unconditional rotation of the
$i$-th qubit, the pulse must be centered at the frequency
$\omega_{qi}^0+2\bar n\mychi$ and have a frequency-width large compared
with $2\sqrt{\bar n}\,\mychi$. For $|\alpha|\geq 1/\pi$ the duration of such a
$\pi$-pulse is thus $T_{\pi}\ll 1/(2\sqrt{\bar n}\mychi)\leq T$. Similarly, the initial coherent
state of the cavity $\ket{\alpha}$ can be prepared from the vacuum by driving the cavity at the frequency $\omega_c-\mychi_A$ with a
pulse of area $\alpha$ and a 
frequency-width large compared with $2N\mychi$; the maximal frequency spread of a cavity dispersively coupled with
strength $\mychi$
to $N$ qubits (Fig.~\ref{fig:1} (b)). Again the duration $T_d$ of this pulse is short since
$T_d\ll 1/(2N\mychi)<T$. The
total duration of the encoding is thus dominated by the dispersive
evolution time $T=\pi/(2\mychi)$, which is independent of $N$ and $M$.
\begin{figure}[ht]
\includegraphics[width=\columnwidth]{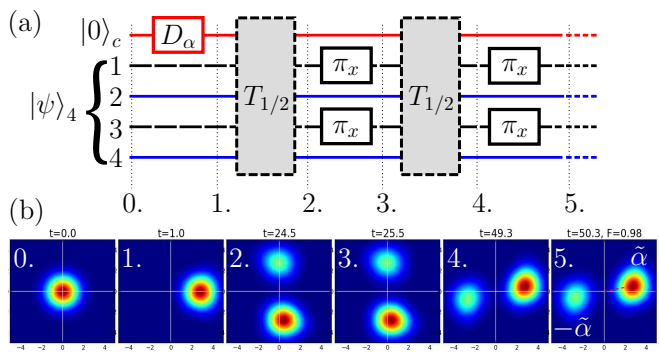}\caption{\label{fig:2}(Color
  online) {\bf (a)} Quantum circuit diagram for encoding the 
  parity of qubits $2$ and $4$ (full (blue) lines). $D_{\alpha}$
  represents the displacement operation, $\pi_x$ a single-qubit
  $\pi$-pulse and $T_{1/2}$ a free dispersive evolution of duration $\pi/(4\chi)$. {\bf (b)}
  Numerical simulation of the evolution of the Q-function of the
  cavity~\cite{Walls_Milburn-2008}, with an initial qubit state $\ket{\psi}_4=(\ket{ggge}+\ket{ggeg}+\ket{eeeg})/\sqrt{3}$. Dissipation from photon loss at a
 rate $\kappa/(2\pi)=10\,{\rm KHz}$ and qubit decoherence with
 $T_1=T_2=20\,{\rm \mu s}$ are included as well as a
  finite displacement and $\pi$-pulse duration of $1\,{\rm ns}$. Other parameters are:
  $\alpha=2$, $\chi/(2\pi)=5\,{\rm MHz}$ and $K/(2\pi)=80\,{\rm KHz}$.
 The self-Kerr term leads to an additional
 phase rotation $\Delta\phi_{\rm nl}=2K\bar n\Delta t$, where $\Delta
 t=50.3\,{\rm ns}$ is the total duration of the encoding. Taking this rotation into account, we obtain a
 fidelity of $F=98\%$ to the
 ideal target state $\ket{\Psi}_{\rm ideal}=\ket{\tilde\alpha_2^{\ }}{\bm
   P}^+_{\{2,4\}}\ket{\psi}_4+\ket{-\tilde\alpha_2^{\ }}{\bm P}^-_{\{2,4\}}\ket{\psi}_4$.} 
\end{figure}

As an example, Fig.~\ref{fig:2} shows the results of a numerical
simulation encoding the parity of $M=2$ out of $N=4$ qubits, which accounts for finite (square) pulse duration, decoherence and qubit-induced cavity
nonlinearity. For the parameter values given in the caption, we find a fidelity
(overlap with the ideal target state) of
$98\%$. By applying single-qubit
rotations to individual qubits before and after the encoding one may
similarly encode the parity of an arbitrary weight $M$ Pauli operator
${\bm Q}_{S_M^{\ }}=\bigotimes_{i\in S_M^{\ }}{\bm\tau_i}$, with
$\bm\tau_i\in\{{\bm\sigma}_i^x,{\bm\sigma}_i^y,{\bm\sigma}_i^z\}$.

{\em Parity readout}. The encoded state is of the form
$\ket{\Psi}_T=\ket{\tilde\alpha_M^{\ }}{\bm
  P}_{S_M}^+\ket{\psi}_N+\ket{-\tilde\alpha_M^{\ }}{\bm
  P}_{S_M}^-\ket{\psi}_N$, where ${\bm P}_{S_M}^{\pm}=(\openone\pm
{\bm Q}_{S_M})/2$ and $\tilde\alpha_M^{\
}=(-i)^Me^{-i\Delta\phi_{\rm nl}}\alpha$. The overlap
  between the two cavity states,
  $\braket{\tilde\alpha_M^{\ }|-\tilde\alpha_M^{\ }}=\exp(-2|\alpha|^2)$, is independent
  of $K$ and $M$. For large $|\alpha|$, these two states are
  distinguishable in principle and a measurement of
 the cavity state is equivalent to a multi-qubit parity measurement. A
 fast readout of the cavity state with $T_{\rm meas}\ll 1/\kappa$,
 may be achieved by lowering the $Q$-factor of the cavity containing
 the qubits ($\kappa\rightarrow\kappa'\gg\kappa$), as recently
 demonstrated~\cite{Yin-2012a_notit}. This $Q$-switching adversely affects the lifetime of the
 qubits via the Purcell effect. However, the latter is expected to be
 weak as long as $\kappa'\ll\mychi$.  Alternatively, the cavity state can be mapped onto the
ancilla qubit, which can subsequently be measured through standard
homodyne measurement of the low-$Q$ readout cavity. The mapping is
achieved physically in three
steps. First, the high-$Q$ cavity field is displaced unconditionally
by $\tilde\alpha_M^{\ }$. To a good
approximation, this maps the encoded state onto
\begin{equation}\label{eq:7}
{\bm D}_{\tilde\alpha_M^{\ }}\ket{\Psi}_T=\ket{2\tilde\alpha_M^{\ }}{\bm
  P}_{{S_M}}^+\ket{\psi}_N+\ket{0}{\bm P}_{{S_M}}^-\ket{\psi}_N\,.
\end{equation}
The second step consists in performing a $\pi$-pulse
on the ancilla qubit, which so far was in its ground state, conditioned on the cavity being in the vacuum
state. As first proposed in~\cite{Leghtas-2013a} and demonstrated in~\cite{Kirchmair-2013a}, this
can be achieved by applying a pulse centered on the bare ancilla qubit
transition frequency, which is narrow in frequency compared with
$8\bar n\mychi_A^{\ }$ (the additional factor of $4$ is due to the
twice as large amplitude of the cavity state in the first term on the
rhs of Eq.~(\ref{eq:7})). Because $\mychi_A^{\ }\gg N\mychi$, a
pulse duration $T_A$ can be chosen such that $1/(2\mychi_A)\ll T_A\ll
1/(2N\mychi)$. For $\bar n>1/4$ the first inequality guarantees
conditionality while the second one allows us to neglect the dispersive
evolution during this operation. The state then becomes approximately
\begin{equation}
\ket{2\tilde\alpha_M^{\ }}\Big({\bm P}_{S_M}^+\ket{\psi}_N\Big)\ket{g}_A+\ket{0}\Big({\bm P}_{S_M}^-\ket{\psi}_N\Big)\ket{e}_A.
\end{equation}
In the third and final step, a displacement of $-2\tilde\alpha_M^{\ }$ is performed
on the cavity conditioned on the ancilla qubit being in the ground
state. This is achieved with a pulse centered at frequency $\omega_c-\mychi_A$, with a frequency-width small compared with
$2\mychi_A^{\ }$ but large compared with $2N\mychi$. Neglecting again the dispersive evolution during this step,
the state finally becomes
\begin{equation}\label{eq:6}
\ket{0}\left[\Big({\bm P}_{S_M}^+\ket{\psi}_N\Big)\ket{g}_A+\Big({\bm P}_{S_M}^-\ket{\psi}_N\Big)\ket{e}_A\right].
\end{equation}
Note that the state~(\ref{eq:6}) is now stationary with respect to the
dispersive interaction, there being no photons in the cavity. Reading
out the state of the ancilla qubit amounts to measuring ${\bm Q}_{S_M}$. After the measurement, the ancilla qubit
may be reset to the ground state efficiently via the readout cavity~\cite{Geerlings-2012a_notit}.

{\em Simulated time evolution}. The ancilla qubit can also be used to simulate the ``time
evolution'' $\exp[-i\theta {\bm Q}_{S_M}]$ under the action of an
arbitrary Pauli operator ${\bm
  Q}_{S_M}$. This is useful e.g.\ for manipulating a logical qubit
state encoded in a stabilizer subspace in which case ${\bm
  Q}_{S_M}$ is taken to be a logical qubit Pauli operator. Starting
from the state Eq.~(\ref{eq:7}), this is achieved
by adiabatically driving the ancilla around a closed loop on its Bloch
sphere subtending
a solid angle $4\theta$, conditioned on there being no photons in the
cavity. The component
in Eq.~(\ref{eq:7}) with zero photons then acquires a
phase of $2\theta$ (half the solid angle) and the state becomes
\begin{align}\label{eq:8}
\ket{2\tilde\alpha_M^{\ }}{\bm
  P}^+_{S_M}\ket{\psi}_N+e^{2i\theta}\ket{0}{\bm
  P}^-_{S_M}\ket{\psi}_N.
\end{align}
Note that $\mychi_A\gg N\mychi$ guarantees that
the adiabatic condition wrt $\mychi_A$ can be satisfied while still
remaining fast wrt the dispersive time scale $1/(N\mychi)$. In (\ref{eq:8}) we omitted the state of the ancilla qubit, since it starts and ends in the
ground state and thus factors out.
The cavity is disentangled from the state by applying the
encoding protocol with $\alpha$ replaced by $-\tilde\alpha_M^{\ }$. After unconditionally
displacing the cavity back to the vacuum, taking into account an
additional nonlinear phase acquired during the decoding, the $N$-qubit
state finally becomes 
\begin{align}
{\bm
  P}^+_{S_M}\ket{\psi}_N+e^{2i\theta}{\bm
  P}^-_{S_M}\ket{\psi}_N=e^{i\theta}e^{-i\theta {\bm Q}_{S_M}}\ket{\psi}_N,
\end{align}
which up to an unimportant global phase factor, represents the action
of the desired unitary.

{\em Application}. 
To test the feasibility of the above protocols, we simulated the
preparation of a logical qubit state of the four-qubit erasure channel
code~\cite{Grassl-1997a}. The stabilizer generators of this code are
$S=\{Z_1Z_2,Z_3Z_4,X_1X_2X_3X_4\}$, where we have switched to the
standard notation $X_i$ and $Z_i$ for the Pauli operators of qubit $i$. The code-space is spanned by the
two $+1$ eigenstates of the stabilizer operators:
\begin{align}
\ket{\pm}&=\frac{1}{2}(\ket{gg}\pm\ket{ee})(\ket{gg}\pm\ket{ee}).
\end{align}
The logical qubit Pauli operators are $\overline Z= X_1X_2=X_3X_4$ and
$\overline X=Z_1Z_3=Z_2Z_4=Z_1Z_4=Z_2Z_3$. Because of the redundancy
of the logical operators, this code protects a logical qubit state $\alpha\ket{+}+\beta\ket{-}$ from the loss (i.e.\ arbitrary error) of a {\em known}
qubit~\cite{Grassl-1997a}.
Here we prepare the logical qubit state
$\ket{\overline\psi}=\exp[-i(\pi/8)\overline X]\ket{+}$ as
follows. $(i)$ We start with the fully polarized four-qubit state
$\ket{gggg}$, which is already a $+1$ eigenstate of the $Z$
stabilizer operators. $(ii)$ Using the encoding protocol, we
measure the logical $\overline Z$
operator $X_1X_2$.
If we obtain $-1$, we apply $Z_1$. $(iii)$ We measure the operator
$X_3X_4$. If we obtain $-1$ we apply
$Z_3$.
$(iv)$ We reset the cavity to
the vacuum. These four steps prepare the
logical state $\ket{+}$. We next use the ancilla
(with $\mychi_A^{\ }=10N\mychi$) to
implement the rotation as described above with ${\bm
  Q}_{S_4}=\overline X$ and $\theta =\pi/8$.
\begin{figure}[ht]
\includegraphics[width=0.9\columnwidth]{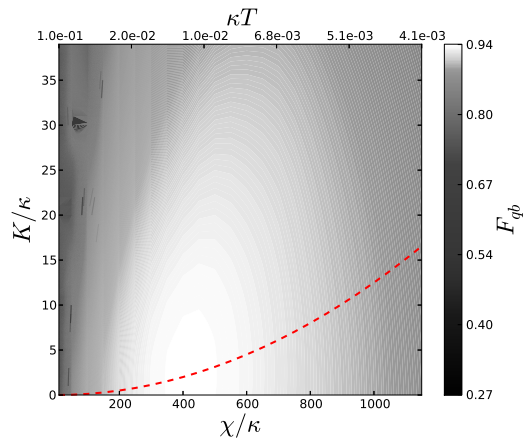}\caption{(Color
  online) Fidelity of the prepared state to
  $\exp[-i(\pi/8)\overline X]\ket{+}$. The (red) dashed
  curve represents the boundary of the inequality $K\geq\chi^2/(4\alpha_q)$ which relates
  the self-Kerr $K$ with the dispersive shift $\chi$ and the
  single-qubit anharmonicity $\alpha_q$~\cite{Nigg-2012a}. Here $\alpha_q/(2\pi)=200\,{\rm
    MHz}$.\label{fig:3}} 
\end{figure}
Fig.~\ref{fig:3} shows
the obtained fidelity
to the ideal target state $\exp[-i(\pi/8)\overline X]\ket{+}$ as a function of $\mychi$ and $K$
in units of $\kappa/(2\pi)=10\,{\rm KHz}$ and for $T_1=T_2=20\,{\mu s}$.
The total duration $T$ of the state preparation is shown on the upper
$x$-axis in units of $1/\kappa$. Focusing on the line $K=0$, when
$\mychi$ is small $T$ is large and dominated by the dispersive
evolutions and the fidelity
is limited by a combination of qubit decoherence, photon loss and
faulty conditional ancilla rotation. The fidelity
then increases with increasing $\mychi$, which reduces the
preparation time $T$ and hence the effects of decoherence and improves
the fidelity of the conditional ancilla rotation. It reaches a maximum in a
regime where both conditional and
unconditional operations can be performed with high
fidelity. A further increase in $\mychi$ degrades the unconditional
cavity displacement and qubit rotations (a $1\,{\rm ns}$
pulse corresponds to a width of $\sim 160\,{\rm MHz}$) and the fidelity
drops. The cavity nonlinearity $K$ and the dispersive shift $\mychi$ are in
fact not independent, but rather related via the
single-qubit anharmonicity $\alpha_q$, by the
inequality $K\geq\mychi^2/(4\alpha_q)$~\cite{Nigg-2012a}, which is
shown as a dashed (red) curve in Fig.~\ref{fig:3} for
$\alpha_q/(2\pi)=200\,{\rm MHz}$.

In conclusion, we proposed a protocol to measure
  stabilizer operators defined on an arbitrary subset of superconducting qubits in the ultra-strong dispersive
regime of cQED. Challenges for the
future will be to extend the present protocol to carry out multiple
stabilizer measurements in parallel and to make it scalable perhaps by
using multiple cavities as in~\cite{DiVincenzo-2012a_notit} and an
efficient encoding of multiple bits of information onto the photonic
Hilbert space~\cite{Leghtas-2013a}. 

We thank L.~Jiang, M.~Mirrahimi, D.~Poulin, M.~Devoret and
R.~Schoelkopf for discussions. The simulations were coded in Python
using the QuTip library~\cite{Qutip-2012a}. This work was supported by
the Swiss NSF, the NSF (DMR-1004406) and the ARO (W911NF-09-1-0514).

\bibliography{/home/sen/phys/library/bibtex/mybib.bib}

\cleardoublepage
\newpage

\begin{widetext}
\begin{center}{\large\bf Supplemental Material for\\``Stabilizer quantum error correction toolbox for
  superconducting qubits'' }\end{center}
\vspace{0.2cm}
\begin{center}Simon E. Nigg and S.~M.~Girvin\\
{\it Departments of Physics, Yale University, New Haven, CT
  06520, USA}\\(Dated: \today)
\end{center}
\begin{center}\begin{minipage}[t]{0.8\textwidth}
\hspace{0.3cm}\small{In these notes, we derive the effective Hamiltonian of Eq.~(1) in the main
text. We generalize the subset parity encoding to arbitrary dispersive shifts. We calculate finite pulse-width effects for
the unconditional displacement and single-qubit rotations and discuss the
constraints imposed on the drive and system spectrum. We describe a
modification of the parity measurement protocol for stabilizer
pumping. We briefly discuss our treatment of decoherence and present the
state tomography of the encoded logical qubit state of the erasure code. Finally we
conclude with a discussion of possible applications and extensions.}
\end{minipage}
\end{center}


\section{Derivation of the effective model of Eq.~(1)}
In terms of its two conjugate variables ${\bm\varphi}$ and ${\bm n}$,
which represent the superconducting phase difference and the number of
Cooper pairs transferred across the
Josephson junction (JJ), a transmon qubit is conventionally described by the Hamiltonian~\cite{Koch-2007a}
\begin{equation}\label{eq:2s}
H=4E_C({\bm n}-n_g)^2+E_J(1-\cos({\bm\varphi}))=4E_C({\bm
  n}-n_g)^2+\frac{E_J}{2}{\bm \varphi}^2+V_{\rm
  nl}({\bm\varphi})\,.
\end{equation}
In the second equality we have explicitly separated the harmonic
and anharmonic
parts of the Josephson potential $E_J(1-\cos({\bm\varphi}))={\bm\varphi}^2E_J/2+V_{\rm
    nl}({\bm\varphi})$. For $E_J\gg E_C$, the ``phase particle'' essentially undergoes
quasi-harmonic oscillations around a local minimum of the cosine
potential, with rare $2\pi$ quantum phase-slip events, which lead
to a weak (exponentially suppressed in $E_J/E_C$)
dependence of the energy levels on the offset charge
$n_g$~\cite{Koch-2007a}. In the following we will neglect the latter
and set $n_g=0$. It is convenient to diagonalize the harmonic
part by introducing the bare mode operator ${\bm
    a}_0=\sqrt{E_J/(2\hbar\omega_0)}\,{\bm\varphi}+2i\sqrt{E_C/(\hbar\omega_0)}\,{\bm n}$
  and frequency $\hbar\omega_0=\sqrt{8E_JE_C}$. Then expanding the
  non-linear potential to leading order (${\bm\varphi}^4$) one obtains
\begin{equation}
H=\hbar\omega_{0}({\bm a}^{\dagger}_0{\bm
  a}_0+1/2)-\frac{E_C}{12}({\bm a}_0+{\bm a}_0^{\dagger})^4+\mathcal{O}({\bm\varphi}^6)
\end{equation}
Note that the coefficient of a given term in the expansion is smaller
than the previous one by a factor $\sim\sqrt{E_C/E_J}$. We are primarily interested in
low-lying energy levels, for which the wavefunctions are localized in
phase i.e. $\sqrt{\braket{{\bm\varphi}^2}}\ll \pi$. Hence we consider here only
the leading order $\varphi^4$ non-linearity.

The coupling of a transmon with the
  electromagnetic field of a
  resonator is typically described in the dipole
  approximation~\footnote{A derivation which goes beyond this simple
    dipole approximation to include the exact linear normal modes is
    given in~\cite{Nigg-2012a,Bourassa-2012b}.} by a
  term $\sim {\bm n}\cdot{\bm E}$ where ${\bm E}=\sum_k i\epsilon_k({\bm
      b}_k-{\bm b}_k^{\dagger})$ is the quantized electric field
    of the (multi-mode) resonator written in terms of its eigenmodes ${\bm
      b}_k$. In a straightforward generalization of the above, the Hamiltonian of $N_{{\rm qb}}$ transmons coupled to multiple
    harmonic modes can be written in the general form (suppressing from now on the zero-point energies)
\begin{equation}
H = \sum_{i=1}^{N_{{\rm qb}}}\hbar\omega_i^{{\rm qb}}{\bm
  a}_i^{\dagger}{\bm a}_i+\sum_j\hbar\omega^{{\rm c}}_j{\bm
  b}_j^{\dagger}{\bm b}_j+\sum_{ij}g_{ij}({\bm a}_i-{\bm
  a}^{\dagger}_i)({\bm
  b}_j-{\bm b}_j^{\dagger}) + \sum_{i=1}^{N_{{\rm qb}}}V^{(i)}_{\rm
  nl}(\bm\varphi_i)\quad\text{with}\quad g_{ij}=\frac{\epsilon_{j}}{2}\sqrt{\frac{\hbar\omega_i^{{\rm qb}}}{E_C^{(i)}}}
\end{equation}
Here $V_{\rm
  nl}^{(i)}(\bm\varphi_i)=E_J^{(i)}[(1-\cos(\bm\varphi_i))-{\bm\varphi_i}^2/2]$. As
before, it is convenient to diagonalize the harmonic part. As shown in~\cite{Nigg-2012a},
this diagonalization can be achieved from the knowledge
of the $N_{\rm qb}\times N_{\rm qb}$ impedance matrix ${\bm Z}$ of the corresponding
linearized circuit with each one of the $N_{{\rm qb}}$ JJs representing a port.
Denoting the
normal modes by ${\bm c}_p$ and their frequencies by $\omega_p$, one
obtains explicitly
\begin{equation}
  \bm\varphi_i^{(k)}=\phi_0^{-1}\sum_{p=1}^M\frac{{\bm
      Z}_{ik}(\omega_p)}{{\bm Z}_{kk}(\omega_p)}\sqrt{\frac{\hbar}{2}\mathcal{Z}_{kp}^{\rm
      eff}}\,({\bm c}_p+{\bm
    c}_p^{\dagger})\quad\text{with}\quad\mathcal{Z}_{kp}^{\rm
    eff}=\frac{2}{\omega_p{\rm Im}[{\partial_{\omega}Y_k(\omega_p)}]}
\end{equation}
Here $\phi_0=\hbar/(2e)$ is the reduced flux quantum and we have
introduced the admittance at port $k$ defined as $Y_k=1/{\bm Z}_{kk}$.
The choice of the port $k$ is arbitrary and reflects a
particular basis choice. The total number of modes is denoted by
$M$. The mode frequencies $\omega_p$ are given by the zeros of the
imaginary part of the admittance, i.e. ${\rm Im}Y_k(\omega_p)=0$ (note
that in the dissipationless case, the elements of ${\bm Z}$ are purely imaginary).

Keeping the leading order $\varphi^4$ non-linearity and normal-ordering
the latter, the Hamiltonian of
the system then takes the form
\begin{align}\label{eq:9s}
H_4 &= \sum_{p}\omega_p{\bm c}_p^{\dagger}{\bm
  c}_p-\sum_{pp'}\gamma_{pp'}\left(2{\bm c}_p^{\dagger}{\bm c}_{p'}+{\bm
    c}_p^{\dagger}{\bm c}_{p'}^{\dagger}+{\bm c}_p{\bm c}_{p'}\right)\\
&-\sum_{pp'qq'}\beta_{pp'qq'}\left(6{\bm c}_p^{\dagger}{\bm
    c}_{p'}^{\dagger}{\bm c}_{q}{\bm c}_{q'}+4{\bm c}_{p}^{\dagger}{\bm
    c}_{p'}^{\dagger}{\bm c}_{q}^{\dagger}{\bm c}_{q'}+4{\bm
    c}_{p}^{\dagger}{\bm c}_{p'}{\bm c}_{q}{\bm c}_{q'}+{\bm c}_{p}{\bm c}_{p'}{\bm
    c}_{q}{\bm c}_{q'}+{\bm c}_p^{\dagger}{\bm c}_{p'}^{\dagger}{\bm
    c}_q^{\dagger}{\bm c}_{q'}^{\dagger}\right)\nonumber,
\end{align}
with
\begin{equation}\label{eq:3s}
\beta_{pp'qq'}=R_0^{-2}\sqrt{\mathcal{Z}_{kp}^{\rm
    eff}\mathcal{Z}_{kp'}^{\rm eff}\mathcal{Z}_{kq}^{\rm
    eff}\mathcal{Z}_{kq'}^{\rm eff}}\,\sum_{i=1}^{N_{{\rm qb}}}\frac{E_J^{(i)}}{6}\frac{{\bm Z}_{ik}(\omega_p)}{{\bm
    Z}_{kk}(\omega_p)}\frac{{\bm Z}_{ik}(\omega_{p'})}{{\bm
    Z}_{kk}(\omega_{p'})}\frac{{\bm Z}_{ik}(\omega_q)}{{\bm
    Z}_{kk}(\omega_q)}\frac{{\bm Z}_{ik}(\omega_{q'})}{{\bm
    Z}_{kk}(\omega_{q'})}
\end{equation}
where $R_0=\hbar/e^2$ and
$\gamma_{pp'}=6\sum_{q}\beta_{qqpp'}$. Eqs.~(\ref{eq:9s}) and
(\ref{eq:3s}) together with a model for the impedance matrix, allow one
to design the circuit Hamiltonian~\cite{Nigg-2012a,Bourassa-2012b}. Importantly, the
above derivation provides the functional dependence among the parameters
$\omega_p$, $\gamma_{ṕp'}$ and $\beta_{pp'qq'}$. By applying a generalized RWA in the dispersive limit of large
detuning between the modes~\footnote{We also neglect multi-photon
  processes.}, we may neglect rapidly-rotating and
non-diagonal terms and thus consider the simplified form
\begin{equation}\label{eq:1s}
H_{\rm eff} = \sum_p\omega_p'{\bm c}_p^{\dagger}{\bm
  c}_p+\frac{1}{2}\sum_{pq}\mychi_{pq}{\bm c}_p^{\dagger}{\bm
  c}_{q}^{\dagger}{\bm c}_q{\bm c}_{p}
\end{equation}
where we have defined $\mychi_{pq}=-24\beta_{pqqp}$ for $q\not=p$ and
$\mychi_{pp}=-12\beta_{pppp}$. The frequencies have been renormalized as
$\omega_p'=\omega_p-2\gamma_{pp}$. The second term on the rhs
of~(\ref{eq:1s}) describes self-Kerr ($\mychi_{pp}$) and cross-Kerr
($\mychi_{qp}$ with $q\not=p$) interactions between
the modes. From the Cauchy-Schwarz inequality and Eq.~(\ref{eq:3s}), it
can be shown that $\mychi_{qp}^2\leq 4\mychi_{qq}\mychi_{pp}$;
equality holding in the single-qubit case. In the dispersive regime of interest here, the
hybridization of qubit and cavity modes is small and one can
unambiguously identify the qubit-like modes as those with the
strongest anharmonicities. Assuming the latter are much larger than
the linewidth and as long as the drive frequencies used are different
from transitions between the first and second excited states, the
qubit Hilbert space can be truncated to the lowest two
levels. From~(\ref{eq:1s}) we then obtain, denoting the cavity modes
with ${\bm a}_i$ and suppressing constant terms
\begin{equation}\label{eq:4s}
H_0=\sum_{i=1}^{N_{{\rm qb}}}\frac{\omega_i^{{\rm
      qb}}}{2}{\bm\sigma}_i^z+\sum_j\omega_j^{\rm c}{\bm
  a}_j^{\dagger}{\bm a}_j
+\sum_{i=1}^{N_{{\rm qb}}}\sum_j\mychi_{ij}{\bm\sigma}_i^z{\bm a}_j^{\dagger}{\bm
  a}_j -\sum_{i}K_i{\bm a}_i^{\dagger}{\bm a}_i^{\dagger}{\bm a}_i{\bm a}_i\,,
\end{equation}
where $K_i\equiv\mychi_{ii}$ and we have kept only the
dispersive and qubit-cavity cross-Kerr shifts. We note that in addition to these
interaction terms there exist in general also qubit-qubit
($\sim{\bm\sigma}_i^z{\bm\sigma}_j^z$, $i\not=j$) and
cavity-cavity ($\sim{\bm a}_i^{\dagger}{\bm a}_i{\bm a}_j^{\dagger}{\bm
  a}_j$, $i\not=j$) cross-Kerr shifts. However in the regime of interest
here, the coefficients of these terms are small and will consequently
be neglected.

A qubit-induced cavity non-linearity
has previously been derived within the dispersive approximation of a
multi-level generalization of the Tavis-Cummings
model~\cite{Boissonneault-2010a}. In addition to the qubit independent term discussed
here, such a treatment also yields qubit dependent non-linearities. In our
derivation, qubit dependent self-Kerr terms appear only at order
$\varphi^6$ and higher and are parametrically small for large ratios $E_J/E_C$. 

We can now apply~(\ref{eq:4s}) to our system depicted in Fig.~1 of the main
text. From the above derivation, we see that in principle all qubits
($N$ active qubits and $1$ ancilla qubit) couple to multiple modes of
both cavities. However, in the weakly coupled dispersive limit for the
readout cavity and
appropriately chosen parameters, we may neglect the coupling of the
active qubits with the readout (low-$Q$) cavity mode and higher modes of the
active (high-$Q$) cavity. Similarly the ancilla approximately only couples to a
single mode in each cavity. Denoting the dispersive shift of the
$N$ qubits by $\mychi_i$ and that of the ancilla with the active cavity mode ${\bm a}$ by $\mychi_A\gg N\mychi$ and with
the readout cavity mode ${\bm b}$ by $\mychi_A'$ we obtain
\begin{equation}
H_0 =\sum_{i=1}^{N}\frac{\omega_i^{{\rm
      qb}}}{2}{\bm\sigma}_i^z+\omega_{\rm c}{\bm
  a}^{\dagger}{\bm a}
+\sum_{i=1}^{N}\mychi_i{\bm\sigma}_i^z{\bm a}^{\dagger}{\bm
  a}+\mychi_A{\bm\sigma}_z^A{\bm a}^{\dagger}{\bm
  a}-K{\bm
  a}^{\dagger}{\bm a}^{\dagger}{\bm a}{\bm a}+H_{\rm readout}\,,
\end{equation}
with $H_{\rm readout}=\omega_{\rm c}'{\bm b}^{\dagger}{\bm b}+\mychi_A'\sigma_z^A{\bm b}^{\dagger}{\bm b}-K'{\bm b}^{\dagger}{\bm
  b}^{\dagger}{\bm b}{\bm b}$. Finally, moving to a frame rotating
with qubits and cavity by the unitary transformation ${\bm U}(t) =
\exp\left[-it\left(\sum_i(\omega_i^{\rm
      qb}/2){\bm\sigma}_i^z+(\omega_c-\mychi_A){\bm a}^{\dagger}{\bm
      a}\right)\right]$ yields:
\begin{equation}\label{eq:10s}
H_0=\sum_{i=1}^N\mychi_i{\bm\sigma}_i^z{\bm a}^{\dagger}{\bm a}-K{\bm
  a}^{\dagger}{\bm a}^{\dagger}{\bm a}{\bm a}\,,
\end{equation}
which for equal dispersive shifts $\mychi_i=\mychi$ reduces to Eq.~(1) of the main
text. This model is a valid when the
ancilla qubit and readout cavity remain in their ground state, in
which case $H_{\rm readout}=0$.

\section{Finite pulse-width effects on unconditional operations}
Having motivated the form of our Hamiltonian~(\ref{eq:10s}) [Eq.~(1) of the main
text], we here use it to investigate the
effects due to finite pulse-width when driving the system with
simple realistic pulses. To
distinguish these effects from the
nonlinear effects we here set $K=0$. In the following
discussion, we concentrate on the unconditional operations necessary
for the parity encoding protocol and neglect the
ancilla qubit and readout cavity. Pulse duration constraints for
the conditional pulses are discussed in~\cite{Leghtas-2013a}. We further assume
independent drives for the qubits and the cavity mode. For the latter
to be a valid approximation, the qubits need to be sufficiently detuned from each
other and from the cavity. In the rotating
frame of Eq.~(\ref{eq:10s}), a cavity drive
at frequency $\omega_d$
is then described by
\begin{equation}
H_d^{\rm c}(t)=\varepsilon_0(t)(e^{it(\omega_{\rm c}-\mychi_A-\omega_d)}{\bm a}^{\dagger}+h.c.).
\end{equation}
In the same rotating frame, a simultaneous drive of the qubits in a
set $S\in S_N=\{1,\dots,N\}$ at frequencies $\omega_i^{d}$ for $i\in S$, is described by a term
\begin{equation}
H_d^{\rm qb}(t)=\sum_{i\in
  S}\frac{\Omega_i(t)}{2}({\bm\sigma}_i^+e^{it(\omega_i^{\rm qb}-\omega_i^{d})}+h.c.).
\end{equation}

\subsection{Unconditional displacement}
At the start of the encoding, we need to be able to displace the
cavity from the vacuum into a coherent state with amplitude $\alpha$
independently of the state of the qubits. We choose $\omega_d
=\omega_c-\mychi_A$ and consider the
Hamiltonian
\begin{equation}\label{eq:5s}
H(t)=\mychi\left(\sum_{i=1}^N\sigma_i^z\right){\bm a}^{\dagger}{\bm a}+\varepsilon_0(t)({\bm
  a}+{\bm a}^{\dagger}).
\end{equation}
Let us denote the computational basis states of $N$ qubits by
$\ket{i_1,\dots,i_N}$, with ${\bm\sigma}_j^z\ket{i_j}=i_j\ket{i_j}$
and $i_j\in\{-1,1\}$. Under the action of the Hamiltonian~(\ref{eq:5s}), the initial product state
$\ket{0}_c\ket{\psi}_N$ with
$\ket{\psi}_N=\sum_{i_1,\dots,i_N}c_{i_1,\dots,
  i_N}\ket{i_1,\dots,i_N}$ evolves into
\begin{equation}\label{eq:6s}
\ket{\psi(t)}=\sum_{i_1,\dots, i_N}c_{i_1,\dots,i_N}\ket{-i\int_0^td\tau\varepsilon_0(\tau)e^{i\mychi(\tau-t)A_{i_1,\dots,i_N}}}\ket{i_1,\dots,i_N},
\end{equation}
where $A_{i_1,\dots,i_N}=\sum_{j=1}^Ni_j$.

For simplicity, we consider a square pulse with envelope
$\varepsilon_0(t)=\epsilon_0\theta(T-t)$. The
amplitude of the coherent state associated with the qubit component
$\ket{i_1,\dots,i_N}$ at time $T$ becomes
\begin{equation}
\alpha_T=-i\epsilon_0Te^{-\frac{i\mychi
      A_{i_1,\dots,i_N}T}{2}}{\rm sinc}\left(\mychi
        A_{i_1,\dots,i_N}T/2\right),
\end{equation}
where ${\rm sinc}(x)=\sin(x)/x$. Thus there is both a qubit-state dependent
rotation and a suppression of the amplitude of the coherent
state. Since $-N\leq
A_{i_1,\dots,i_N}\leq N$, if $N\mychi T\ll 1$, we can
expand and obtain
\begin{equation}
\alpha_T = -i\epsilon_0T - \epsilon_0 T^2\mychi
\frac{A_{i_1,\dots,i_N}}{2}+\mathcal{O}((N\mychi T)^2).
\end{equation}
Thus the first-order correction to the displacement of the cavity
state is a qubit-state dependent rotation. The change in amplitude
is second order in $N\mychi T$. This has important consequences for the
encoding protocol, since it allows for a partial compensation
of the finite pulse-length effect of the displacement simply by shortening the duration of the
subsequent free evolution part. Retaining only the leading order term, the state in (\ref{eq:6s})
factorizes and we have
\begin{equation}\label{eq:7s}
\ket{\psi(T)}\approx \ket{-i\epsilon_0T}\ket{\psi}_N.
\end{equation}
which corresponds to a cavity displaced by $\alpha=-i\epsilon_0T$,
while the qubits were left unchanged. In terms of the drive strength
and final coherent state amplitude,
this is valid when $|\alpha|\ll \epsilon_0/(N\mychi)$. Because $-N\leq
A_{i_1,\dots,i_N}\leq N$, the maximum amplitude difference is given to
first order by
$\epsilon_0T^2N\mychi=|\alpha|N\mychi T$. We may then define a fidelity of
the displacement operation as the absolute value of the overlap
between an ideally displaced coherent state and a state with maximal
phase difference. To leading order, this is found to be
\begin{equation}
F\approx \exp\left[-\frac{1}{8}\left(|\alpha| T\mychi\right)^2\right]\,.
\end{equation}
Thus we see that to achieve high fidelity, the average number of photons in
the cavity $\bar n=|\alpha|^2$ should satisfy $\sqrt{\bar n}N\mychi T\ll
8$. Fig.~\ref{fig:s1} shows the numerically computed fidelity to the
target displaced state obtained from the initial state
$\ket{0}_c\ket{gggg}$ as a function of the square pulse duration
$T$. For this simulation we took $\mychi/(2\pi)=10\,{\rm
  MHz}$, $\alpha=2.5$.
\begin{figure}[ht]
{
\includegraphics[width=0.6\textwidth]{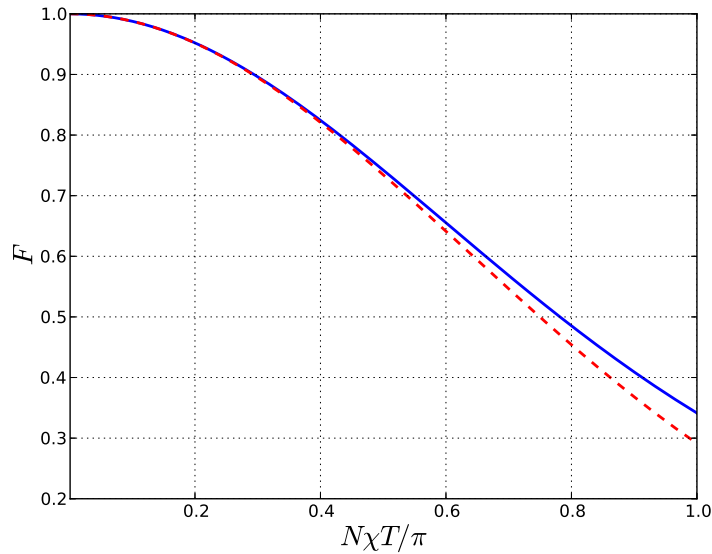}}\caption{Fidelity of un-conditional displacement for the initial state
  $\ket{0}_c\ket{gggg}_{N=4}$ as a function of square pulse duration
  $T$. The dashed (red) curve shows
  $F=\exp[-(|\alpha|N\chi T)^2/8]$. We took a dispersive shift $\chi/(2\pi)=10\,{\rm
  MHz}$, and a cavity displacement $\alpha=2.5$.\label{fig:s1}}
\end{figure}


\subsection{Unconditional qubit selective $\pi$-pulses}
In order to perform subset-selective parity measurements, one needs to perform
unconditional $\pi$-pulses on subsets of qubits as explained in the
main text. Following the discussion in the main text we choose the
drive frequencies to be $\omega_i^d=\omega_i^{\rm qb}+2\bar n\mychi$
and thus consider the following Hamiltonian in the rotating frame
\begin{equation}
H(t) =\mychi\left(\sum_{i=1}^N\sigma_i^z\right){\bm a}^{\dagger}{\bm a}+\sum_{j\in
  S_M^c}\frac{\Omega_{j}(t)}{2}\left(e^{2i\bar n\mychi t}{\bm\sigma}_j^-+h.c.\right).
\end{equation}
Note that here we have neglected the off-resonant driving of the
qubits in $S_M$. In order for this to be valid, we must require the
frequency detuning between any two qubits to be large compared
with $2\sqrt{\bar n}\mychi$, such that we may find a pulse
  duration $T$ satisfying $|\omega_{qi}-\omega_{qj}|_{i\not=j}\gg T^{-1}\gg
  2\sqrt{\bar n}\mychi$. Moving to a frame rotating with the qubits in $S_M^c$ at frequency
$\bar n\mychi$, we can write the Hamiltonian as (we use the same symbol
$H$ to denote the Hamiltonian in the new frame)
\begin{equation}\label{eq:8s}
H=\mychi\left(\sum_{j\in S_M}{\bm\sigma}_j^z\right){\bm a}^{\dagger}{\bm
  a}+\mychi\left(\sum_{j\in S^c_M}{\bm\sigma}_j^z\right)({\bm
  a}^{\dagger}{\bm a}-\bar
n)+\sum_{j\in S_M^c}\frac{\Omega_{j}(t)}{2}{\bm \sigma}_j^x\,.
\end{equation}
Clearly the first term on the rhs commutes with the remaining terms and
leads to a ``free'' evolution of the qubits in $S_M$ and of the
cavity given by ${\bm U}_{S_M}(t)=\exp(-it\mychi{\bm a}^{\dagger}{\bm
  a}\sum_{i\in S_M}{\bm\sigma_i^z})$. The action of the last two terms on each
qubit in $S_M^c$ is described by a single qubit
Hamiltonian of the form
\begin{equation}\label{eq:19s}
H_1=\mychi{\bm\sigma}^z({\bm a}^{\dagger}{\bm a}-\bar
n)+\frac{\Omega(t)}{2}{\bm \sigma}^x.
\end{equation}
Defining $\beta(t)=\int_0^t\frac{\Omega(\tau)}{2}d\tau$, the associated evolution operator is
\begin{equation}
{\bm U}_1(t)=e^{-i\beta(t){\bm\sigma}^x}e^{-i\mychi ({\bm
    a}^{\dagger}{\bm a}-\bar n)\int_{0}^td\tau\left\{{\bm\sigma}^z[\cos^2(\beta(\tau))-\sin^2(\beta(\tau))]+2{\bm\sigma}^y\sin(\beta(\tau))\cos(\beta(\tau))\right\}}.
\end{equation}
For $\mychi=0$ this reduces to a simple ${\bm\sigma}^x$ rotation,
the angle of which is determined by the integral of the pulse
envelope. The total evolution operator, including the contribution from
the first term on the rhs of~(\ref{eq:8s}), is then
\begin{equation}
{\bm U}(t)={\bm U}_{S_M}(t)e^{-i\beta(t)\sum_{j\in S_M^c}{\bm\sigma}_j^x}e^{-i\mychi
  ({\bm a}^{\dagger}{\bm a}-\bar n)\sum_{j\in S_M^c}\int_{0}^td\tau\left\{{\bm\sigma}_j^z[\cos^2(\beta(\tau))-\sin^2(\beta(\tau))]+2{\bm\sigma}_j^y\sin(\beta(\tau))\cos(\beta(\tau))\right\}}.
\end{equation}

For a square pulse envelope $\Omega(t)=\Omega_0\theta(T-t)$ with
$T=\pi/\Omega_0$ such that $\beta(T)=\pi/2$ it is easy to
show that
\begin{equation}
{\bm U}(T)={\bm U}_{S_M}\left(\frac{\pi}{\Omega_0}\right)e^{-\frac{i\pi}{2}\sum_{j\in
    S_M^c}{\bm\sigma}_j^x}e^{-\frac{2i\mychi}{\Omega_0}({\bm
    a}^{\dagger}{\bm a}-\bar
  n)\sum_{j\in S_M^c}{\bm\sigma}_j^y}.
\end{equation}
We are interested in the action of this operator on a coherent state or superposition of coherent states, for which the photon number is Poisson
distributed around $\bar n$ with variance equal to $\bar n$. Hence if
$2\sqrt{\bar n}\mychi\ll \Omega_0$, we have
\begin{equation}
{\bm U}(T)\approx {\bm U}_{S_M}(T)\left[(-i)^M\prod_{j\in
    S_M}{\bm \sigma}_j^x+\mathcal{O}\left(\frac{2\sqrt{\bar n}\mychi T}{\pi}\right)\right].
\end{equation}
If furthermore, we require that $2\mychi M \pi\ll \Omega_0$, then ${\bm U}_{S_M}\approx
\openone+\mathcal{O}(2\mychi MT)$ so that
\begin{equation}
{\bm U}(T)\approx (-i)^M\prod_{j\in
  S_M}{\bm\sigma}_j^x+\mathcal{O}\left(\frac{2\sqrt{\bar
      n}\mychi T}{\pi},2\mychi MT\right).
\end{equation}

Fig.~\ref{fig:s2} shows the fidelity as a function of the square pulse duration $T$,
of an unconditional two-qubit $\pi$
rotation acting on an
initial state $\ket{\alpha}_c\ket{\psi}_4$ with
$\ket{\psi}_4=\sqrt{1/3}(\ket{eegg}+\ket{eggg}-\ket{egge})$, $\alpha=2.5$ and for
$\mychi/(2\pi)=10\,{\rm MHz}$.
\begin{figure}[ht]
{
\includegraphics[width=0.6\textwidth]{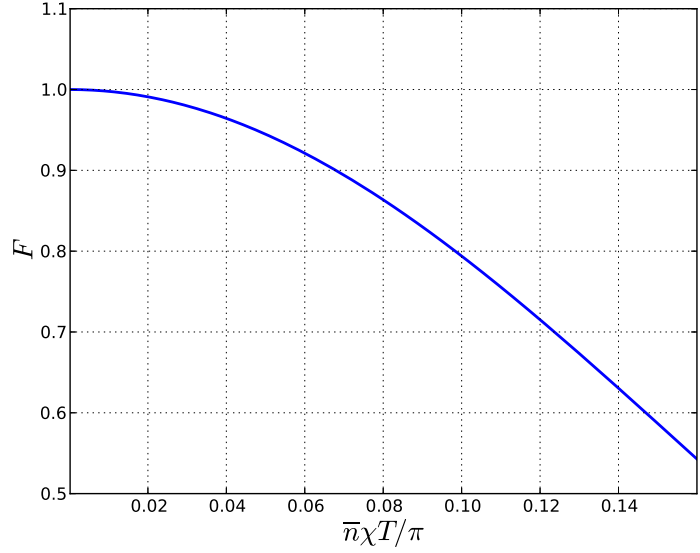}}\caption{Fidelity
  of un-conditional $\pi$ rotation of the first two qubits of the initial state
  $\ket{\alpha}_c\ket{\psi}_4$ as a function of square pulse duration
  $T$. The cavity displacement is $\alpha=2.5$ and the dispersive
  shift is $\chi/(2\pi)=10\,{\rm MHz}$.\label{fig:s2}}
\end{figure}
\subsection{Fidelity of subset selectivity}
The infidelity of the displacement and $\pi$ pulses, due to finite pulse
duration discussed in the previous two sections, reduces the fidelity
of the subset selectivity. To illustrate this, Fig.~\ref{fig:s6} shows the fidelity of
the subset parity encoding of two out of four qubits discussed in the
main text (See Fig.~2 in the main text), as a function of the displacement and
$\pi$-pulse duration. As expected, the fidelity decreases
monotonically with increasing pulse duration. As pointed out above, by
reducing the duration of
the free evolution periods, one can partially compensate for the
finite duration of the displacement operation. This is illustrated by
the full curve in Fig.~\ref{fig:s6}, obtained by subtracting the
displacement pulse duration from the free evolution periods. In contrast, the dashed curve shows the result without the
correction. For the simulation we used square pulses and the same parameters (including
dissipation and Kerr-nonlinearity) as in the main text. It is
natural to expect that more sophisticated pulse shapes designed by
optimal control techniques can be used to significantly
improve the encoding fidelity, by suppressing the spectral weight at
undesired frequencies.
\begin{figure}[ht]
{
\includegraphics[width=0.6\textwidth]{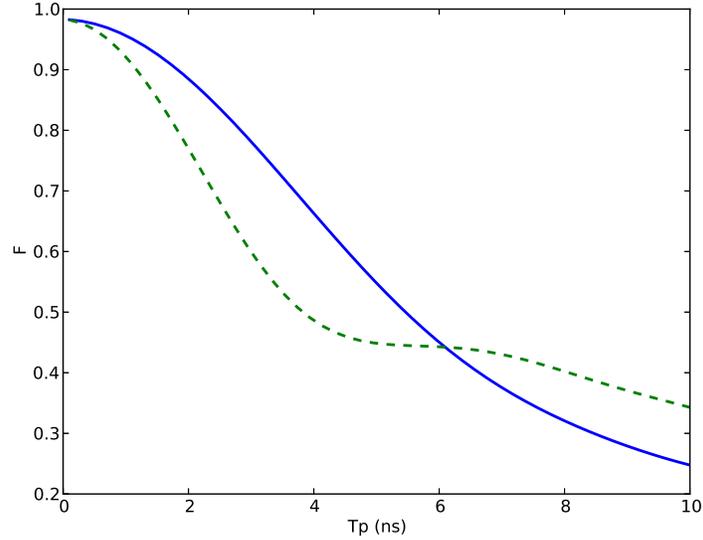}}\caption{Fidelity
  of subset parity encoding as a function of pulse duration. The
  parameters are the same as used in the example given in the main
  text. Full curve: encoding with shortened free evolution
  periods ($T_{\rm free}\rightarrow T_{\rm free}-T_p$). Dashed curve: encoding without adjustment of free
  evolution period.\label{fig:s6}}
\end{figure}
\section{Subset parity encoding for arbitrary dispersive shifts}
In the main text we discuss the case of equal dispersive shifts
$\mychi_i=\mychi$. Although this could in principle be realized by fine tuning of the
coupling strengths and detunings of the qubits with respect to the cavity mode,
we here discuss a simple way to circumvent the need for
tunability. This is advantageous since tunability invariably leads to
shorter coherence times. For variable dispersive shifts, the main difference is that we now
need to apply appropriately timed $\pi$-pulses to {\em all} qubits. For
simplicity we consider here instantaneous pulses and disregard the
non-linear cavity self-Kerr term. Thus we start from the dispersive
interaction
\begin{equation}
H_0 =\left(\sum_{j=1}^N\mychi_j\bm\sigma_j^z\right){\bm a}^{\dagger}{\bm a}.
\end{equation}

\begin{figure}[ht]
\includegraphics[width=0.7\textwidth]{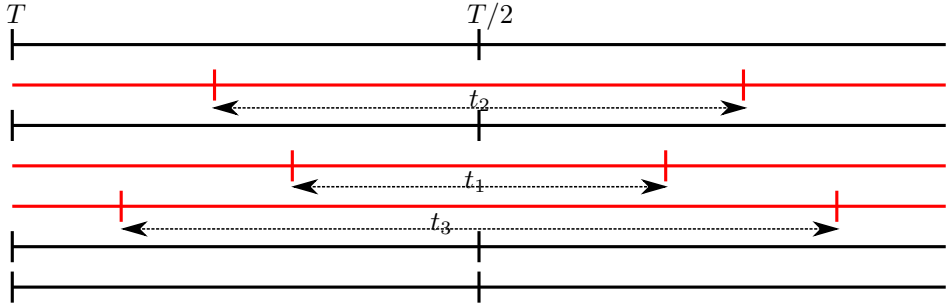}\caption{Schematic
representation of pulse sequence for arbitrary dispersive shifts. Each
horizontal full line represents the time-line of a qubit the time origin
being at the far right end. The vertical
lines indicate the positions in time of bit-flips
(pi-pulses).\label{fig:s1n}}
\end{figure}

As illustrated in Fig.~\ref{fig:s1n}, the idea is to use a pair of bit-flips
for each qubit appropriately separated in time to adjust the cavity-phase contribution of
each individual qubit. To simplify the discussion, let us label the
$M\leq N$ qubits we wish to encode the parity of by the indices
$S_M=\{1,\dots M\}$. Denoting the total duration of the pulse sequence
by $T$, the condition for
qubit $j\in S_M$ to contribute to the encoded parity is
\begin{equation}
\mychi_j(T-2t_j)=\frac{\pi}{2}\Leftrightarrow t_j=\frac{T}{2}-\frac{\pi}{4\mychi_j},
\end{equation}
where $t_j\leq T$ is the time delay between two bit-flips (applications
of ${\bm\sigma}_j^x$). To cancel the contribution to the total parity of the
qubits not in $S_M$ it suffices to require that $t_j=T/2$ for $j\notin S_M$.
Let us now further order the indices $t_j$ for $j\in S_M$ in 
order of increasing magnitude, i.e. $0\leq t_1\leq t_2\leq\dots\leq
t_M\leq T$.

The following operator then implements the desired subset-parity encoding
\begin{equation}
U_T=\left(\prod_{j=M+1}^N{\bm\sigma}_j^x\right)e^{-iH_0\frac{T-t_M}{2}}{\bm\sigma}_M^xe^{-iH_0\frac{t_M-t_{M-1}}{2}}{\bm\sigma}_{M-1}^x\dots{\bm\sigma}_1^xe^{-iH_0\frac{t_1}{2}}\left(\prod_{j=M+1}^N{\bm\sigma}_j^x\right)e^{-iH_0\frac{t_1}{2}}{\bm\sigma}_1^xe^{-iH_0\frac{t_2-t_1}{2}}\dots{\bm\sigma}_M^xe^{-iH_0\frac{T-t_M}{2}}.
\end{equation}
Indeed, after some algebra one finds that
\begin{equation}
U_T=\prod_{j\in S_M}e^{-i\mychi_j(T-2t_j){\bm\sigma}_j^z{\bm a}^{\dagger}{\bm
    a}}=\prod_{j\in S_M}e^{-i\frac{\pi}{2}{\bm\sigma}_j^z{\bm
    a}^{\dagger}{\bm a}}.
\end{equation}
Note that the time ordering used here is not unique and depending
on the experimental conditions, a particular choice may be favored over
others. Acting on a coherent state with amplitude $\alpha$ one finds
\begin{equation}
U_T\ket{\alpha}=\ket{(-i)^M\alpha \prod_{j\in S_M}{\bm\sigma}_j^z},
\end{equation}
which is the desired subset parity mapping. Note that the condition $t_j\geq 0$ implies that $T\geq
\pi/(2\min_{j\in S}(\mychi_j))$. For $\mychi_j=\mychi$ and $T=\pi/(2\mychi)$ we
recover the case discussed in the main text.

\section{Stabilizer pumping}
In the main text we show how to measure a stabilizer by first encoding
its eigenvalues onto the cavity state and then reading out
the latter either directly or via the ancilla qubit and the readout cavity. 
Alternatively one can also implement
conditional jump operators of the form ${\bm\sigma_i^x}{\bm
  P}_{S_M}^{-}$ where ${\bm P}_{S_M}^{\pm}=(\openone\pm{\bm
  Z}_{S_M})/2$ with ${\bm Z}_{S_M}=\prod_{i\in
  S_M}{\bm\sigma}_i^z$ and $i\in S_M$. Such operators can be used to
digitally simulate an artificial bath that pumps the system onto the stabilizer
subspace as first proposed for trapped ions
in~\cite{Barreiro-2011a,Muller-2011a}. The dissipative evolution under
the action of such jump operators can be
implemented with the
protocol shown in Fig.~\ref{fig:s3}. As for a parity measurement, one first
encodes the eigenvalues of the parity operator onto the cavity
state and applies and unconditional displacement to obtain the state
(see Eq.~(5) of the main text)
\begin{equation}
\ket{\Psi}=\ket{2\tilde\alpha_N^{\ }}{\bm
  P}_{{S_M}}^+\ket{\psi}_N+\ket{0}{\bm P}_{{S_M}}^-\ket{\psi}_N\,.
\end{equation}
Instead of mapping the cavity state onto an ancilla as in
the measurement protocol, one then applies a rotation by an angle
$\theta$ around $x$ of qubit $i\in
S_M$ {\em conditioned} on the cavity being
in the vacuum state. Note that this does not require an ancilla qubit. Because $\{\sigma_i^x,Z_{S_M}\}_+=0$ for $i\in
S_M$, the state becomes
\begin{equation}\label{eq:14s}
{\bm
  P}_{S_M}^+\left(\ket{\psi}_N\ket{2\alpha}_c+i\sin(\theta/2)\sigma_i^x\ket{\psi}_N\ket{0}_c\right)+\cos(\theta/2){\bm
  P}_{S_M}^-\ket{\psi}_N\ket{0}_c.
\end{equation}
By resetting the cavity to the vacuum (via an ancilla qubit or
a cavity $Q$-switch~\cite{Yin-2012a_notit}) and repeating the procedure one generates the
dissipative (non-unitary) evolution~\cite{Muller-2011a}
\begin{equation}
\ket{0}\bra{0}\otimes\rho_{N}\rightarrow\ket{0}\bra{0}\otimes\left({\bm
    E}_0(\theta)\rho_{N}{\bm E}_0^{\dagger}(\theta)+{\bm E}_x^{(i)}(\theta)\rho_N{\bm E}_x^{(i)\dagger}(\theta)\right)\,,
\end{equation}
with ${\bm E}_0(\theta)={\bm P}_{S_M}^++\cos(\theta/2){\bm P}_{S_M}^-$
and ${\bm E}^{(i)}_x(\theta)=i\sin(\theta/2){\bm\sigma_i^x}{\bm P}_{S_M}^-$.
Thus with probability $\sin^2(\theta/2)$, a $-1$ eigenstate of ${\bm
  Z}_{S_M}$ is converted into a $+1$ eigenstate. For $\theta=\pi$, this ``stabilizer pumping'' occurs with unit probability. Generalization to arbitrary
weight-$M$ Pauli-operators can be achieved with suitable single
qubit rotations.
\begin{figure}[ht]
\hspace{0.5\textwidth}
\Qcircuit @C=1em @R=.7em {
\lstick{\ket{0}_c}  & \gate{D_{\alpha}} & \multigate{4}{T_{1/2}}&\qw &
\multigate{4}{T_{1/2}}&\qw&\gate{D_{\alpha}}&\ctrl{3}&\gate{{\rm
    reset} \ket{0}_c}&\qw\\
\lstick{1}  & \qw & \ghost{T_W} &\multigate{3}{\Pi_{S_M^c}}&\ghost{T_W}&\qw&\multigate{3}{\Pi_{S_M^c}}&\qw&\qw&\qw\\
\lstick{\vdots}  & \qw & \ghost{T_W} & \ghost{\Pi_{S_M^c}}&\ghost{T_W}&\qw&\ghost{\Pi_{S_M}}&\qw&\qw&\qw\\
\lstick{\vdots}  & \qw & \ghost{T_W} &\ghost{\Pi_{S_M}}&
\ghost{T_W}&\qw&\ghost{\Pi_{S_M}}&\gate{X^0_{i\in S_M,\theta}}&\qw&\qw\\
\lstick{N}  & \qw & \ghost{T_W} &\ghost{\Pi_{S_M}}&\ghost{T_W}&\qw&\ghost{\Pi_{S_M}}&\qw&\qw&\qw
}\caption{Quantum circuit diagram for stabilizer
  pumping. $D_{\alpha}$ denotes an unconditional displacement by
  $\alpha$. $T_{1/2}$ represents a free evolution of duration
  $T_{1/2}=\pi/4\chi$. $\Pi_{S_M^c}$ represents a (parallel)
  multi-qubit $\pi$-pulse on qubits in $S_M^c$. $X^0_{i\in S_M,\theta}$
  represents a conditional rotation of qubit $i$ by $\theta$ around
  the $x$-axis, conditioned on
  the cavity being in the vacuum. ${\rm reset}\ket{0}_c$ represents
  the (dissipative) cavity reset operation.\label{fig:s3}}
\end{figure}
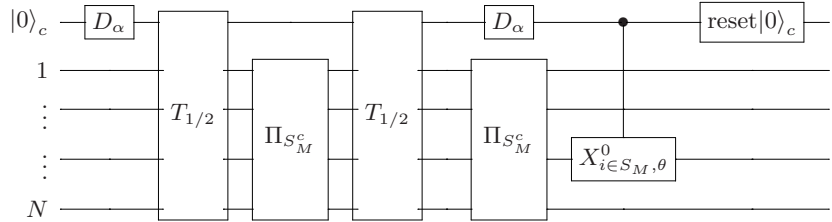

\section{Decoherence}
Decoherence due to photon losses and qubit decay and dephasing are
treated within the Lindblad master equation
formalism~\cite{Breuer-Petruccione}. The master equation we solve numerically is
\begin{equation}
\dot\rho =-i[H,\rho]+\kappa\mathcal{D}[{\bm
  a}]\rho+\sum_i\left(\frac{\Gamma_{\varphi}}{2}\mathcal{D}[{\bm\sigma_i^z}]\rho+\Gamma_-\mathcal{D}[{\bm\sigma}^-_i]\rho+\Gamma_+\mathcal{D}[{\bm\sigma}^+_i]\rho\right)\,,
\end{equation}
where $\mathcal{D}[{\bm L}]\rho = (2{\bm L}\rho{\bm L}^{\dagger}-{\bm
  L}^{\dagger}{\bm L}\rho-\rho{\bm L}^{\dagger}{\bm L})/2$. Here $\kappa$
is the single photon decay rate of the cavity,
$\Gamma_{\varphi}$ is the phase relaxation (pure dephasing) rate of
the qubits and the qubit relaxation rate is given by
$1/T_1=\Gamma_-+\Gamma_+$. We consider the zero-temperature limit $\Gamma_+=0$.
\section{Logical qubit state tomography}
In Fig.~3 of the main text, we show the state fidelity of the encoded logical qubit
state to the target state $\exp[-i(\pi/8)\overline X]\ket{+}$ as a function of
the dispersive shift $\mychi$ and cavity nonlinearity $K$. In the left
panel of Fig.~\ref{fig:s4}, we
compare the resulting four-qubit state with the ideal target state in
the Pauli-bar representation of the reduced density matrix (tracing
out the cavity), for $K/(2\pi)=80\,{\rm KHz}$, $\mychi/(2\pi)=5\,{\rm MHz}$ and
$\kappa/(2\pi) = 10\,{\rm KHz}$. The fidelity to the target state is
$94\%$. The right panel of Fig.~\ref{fig:s4} shows the
Bloch-sphere representation of the prepared logical qubit state. From
the latter, we see that the infidelity is due to both a loss of purity and a deviation
from the ideal rotation.
\begin{figure}[h!t]
\includegraphics[width=0.55\textwidth]{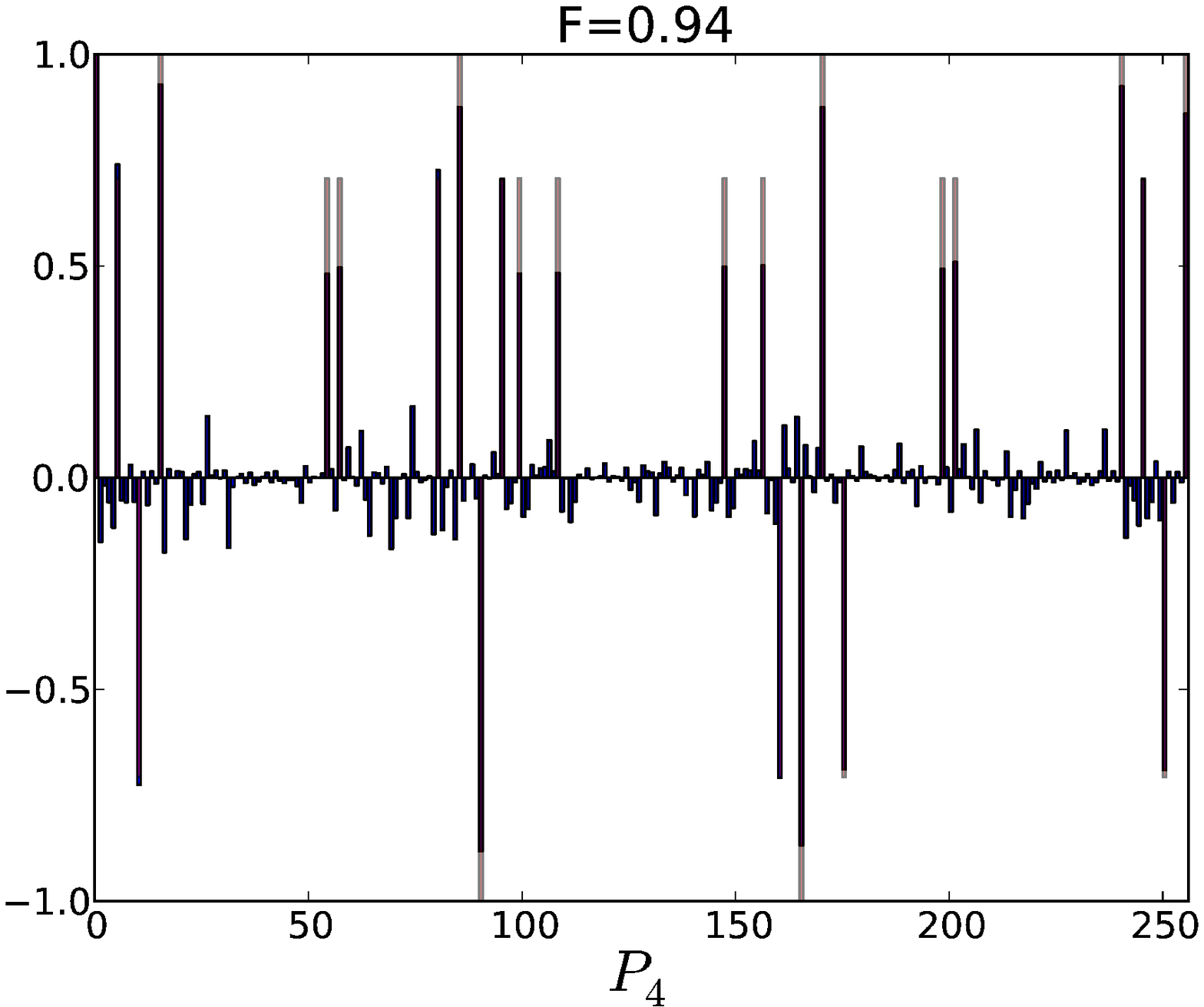}\hfill\includegraphics[width=0.35\textwidth]{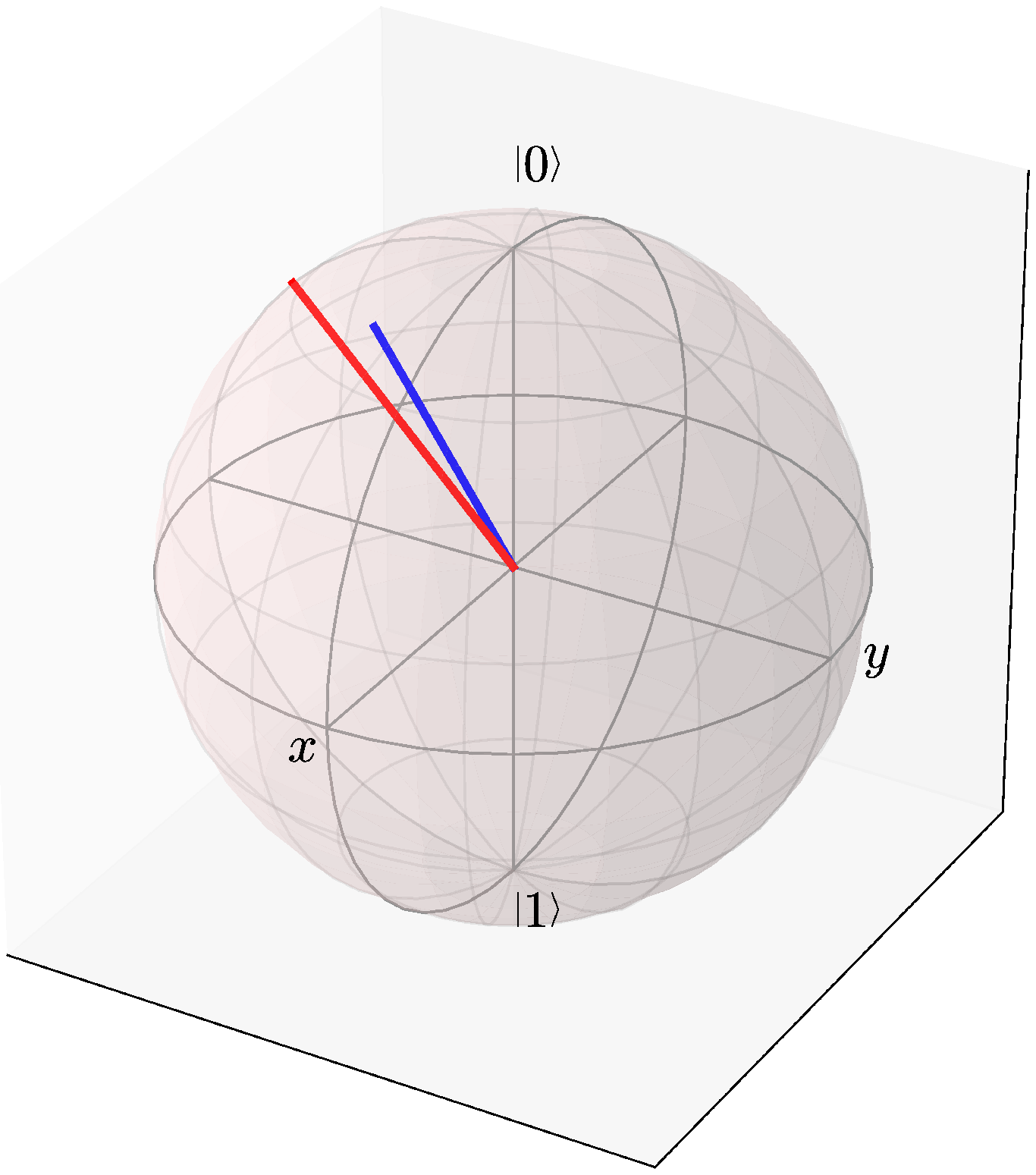}\caption{(Color
  online) {\bf Left panel:} Pauli-bar representation of the prepared
  four qubit state (full
  (blue) bars) and of the target state $\exp[-i(\pi/8)\overline
  X]\ket{+}$ (lighter (red) bars). Here ${\bm
    P}_4=\{\openone,{\bm\sigma}^x,{\bm\sigma}^y,{\bm\sigma}^z\}^{\otimes
    4}$ is the Pauli-group of order four. {\bf Right panel:} Bloch-sphere
  representation of the prepared logical qubit state (full (blue)
  vector) and corresponding target state (full (red) vector).\label{fig:s4}}
\end{figure}

\section{Discussion \& Outlook}
We conclude these notes by an informal discussion of a few possible applications of
our protocol to illustrate its versatility.

\subsection{Implementation of the toric code}
Because we can measure Pauli operators on
arbitrary subsets of qubits, it becomes possible to implement ``in software'' non-trivial
topologies such as required e.g. by the toric
code~\cite{Kitaev-2003a}, in which $2\frak{g}$ logical qubits can be encoded into
the ground state stabilizer subspace of a qubit manifold with genus $\frak{g}$.
Let us consider the case $\frak{g}=1$ of a ring torus as depicted in
Fig.~\ref{fig:s5}.
\begin{figure}
\includegraphics[width=0.7\textwidth]{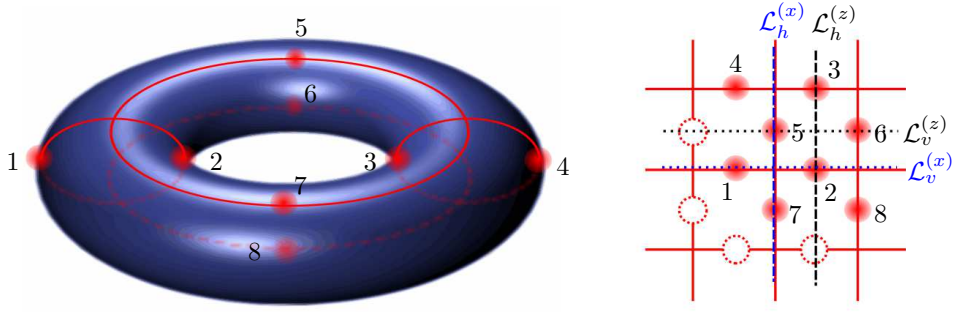}\caption{{\bf Left panel.} $8$-qubit
  toric code. The qubits are represented by read beads. {\bf Right
    panel.} Corresponding planar representation. The latter is easier
  to implement experimentally and the toric boundary conditions are
  imposed when defining the
  stabilizers. $\mathcal{L}_{i}^{(j)}$ indicate the
  incontractible loops used to define the logical qubit operators (see
  text).\label{fig:s5}}
\end{figure}
The stabilizers of the $2N$-qubit toric code for
example are the $N$ plaquette ${\bm P}_i$ and $N$ star operators ${\bm S}_i$
defined as
\begin{equation}
{\bm P}_i=\prod_{j\in {\rm plaquette}_i}{\bm\sigma}_j^z\,,\quad{\bm
  S}_i=\prod_{j\in{\rm star}_i}{\bm\sigma}_j^x\,.
\end{equation}
Crucially, because of the toric boundary conditions it holds true that
$\prod_{i\in{\rm lattice}}{\bm P}_i=\prod_{i\in{\rm lattice}}{\bm
  S}_i=\openone$. These two constraints then lead to a four-fold
degeneracy ($2^{2\frak{g}}$-fold degeneracy in general) of the ground state manifold. A ground state of the toric code can be
prepared by starting from the fully polarized state $\ket{g,\dots,g}$
of the $2N$ qubits in the $Z$ basis and sequentially measuring all the
star operators and applying a corrective single
qubit $Z$ operation on a qubit in any star with
eigenvalue $-1$. Fig.~\ref{fig:s5} illustrates a minimal $8$-qubit toric code and
its planar implementation. The stabilizers of this code are explicitly:
\begin{align}
  {\bm P}_1&=Z_1Z_4Z_7Z_8 & {\bm S}_1=X_1X_2X_5X_7\nonumber\\
  {\bm P}_2&=Z_2Z_3Z_7Z_8 & {\bm S}_2=X_3X_4X_5X_7\nonumber\\
  {\bm P}_3&=Z_1Z_4Z_5Z_6 & {\bm S}_3=X_1X_2X_6X_8\nonumber\\
  {\bm P}_3&=Z_2Z_3Z_5Z_6 & {\bm S}_3=X_3X_4X_6X_8
\end{align}
Here $Z_i={\bm\sigma}_i^z$ and $X_i={\bm\sigma}_i^x$. Clearly,
unfolding the torus onto a plane entails the appearance of non-local
stabilizer operators. Finally we note that our protocol can also be used to measure the logical Pauli operators
of the toric code, which are defined in terms of the incontractible
$X$ or $Z$ loop operators
winding the torus in either
vertical or horizontal directions (see Fig.~\ref{fig:s5}). Explicitly
\begin{align}
\overline{Z}^{(1)}&=\prod_{i\in\mathcal{L}^{(z)}_h}{\bm\sigma}_i^z&\overline{Z}^{(2)}=\prod_{i\in\mathcal{L}^{(z)}_v}{\bm\sigma}_i^z\nonumber\\
\overline{X}^{(1)}&=\prod_{i\in\mathcal{L}^{(x)}_v}{\bm\sigma}_i^x&\overline{X}^{(2)}=\prod_{i\in\mathcal{L}^{(x)}_h}{\bm\sigma}_i^x\,.
\end{align}

\subsection{Preparation of cluster states \& measurement based quantum
computation}
Cluster states are the central resource of measurement based
quantum computation~\cite{Raussendorf-2000a,*Raussendorf-2003a}. A cluster state is a stabilizer state
defined on a graph $G$, where each vertex $i$ is associated with a
stabilizer generator of the form ${\bm S}_i = X_i\prod_{j\in{\rm
    nn}(i)}Z_j$ where ${\rm nn}(i)$ denotes nearest neighbor vertices
to vertex $i$. By applying a single qubit Hadamard rotation
on qubit $i$ at the beginning and at the end, we can use our encoding procedure to measure ${\bm
  S}_i$ and hence prepare arbitrary cluster states. To perform a
quantum computation one requires in addition the ability to perform single-qubit
measurements. Because single
qubit operators are weight-one Pauli operators, we can again use our
measurement protocol to perform such measurements and hence perform
one-way measurement based quantum computation. In practice these
single-qubit measurements can be done either
as explained in the main text, by applying parallel echo $\pi$-pulses on all the
qubits except the one being measured, or alternative by first
performing a parity measurement of all the (appropriately rotated)
qubits (i.e. without any echo $\pi$-pulses) yielding $P_{\rm
  tot}\in\{\pm 1\}$ and then repeating the
measurement, echoing-out only the qubit that is to be measured
yielding $P_{\rm comp}^{(i)}\in\{\pm 1\}$. The parity of the qubit to be
measured is then given by $P_i=P_{\rm tot}/P_{\rm
  comp}^{(i)}$. The advantage of the latter approach is a drastic
reduction in the number of necessary $\pi$-pulses (from $N-1$ to $1$
where $N$ is the total number of qubits). This reduction
applies in general for subset parity measurements when the total number of
qubits $N$ is larger than twice the number of qubits in the subset $M$.
\subsection{Conclusion}
The proposed protocols offer a direct route to implement
stabilizer quantum error correction and to ``digitally simulate `` models
of topological quantum computation with superconducting qubits in the
ultra-strong dispersive regime. From an experimental point of
view, the main challenge presumably lies in engineering a system of multiple qubits
with long coherence times and with sufficient separation of time scales to validate
the approximations made in our derivations. A proof of principle
experiment seems within reach of current technology.

As an outlook for future research, an interesting question is whether
our approach can be extended to measure all stabilizers of a given
code in parallel rather than sequentially. This should be possible in principle
as these operators all commute with each other. A naive extension to
encode an entire syndrome (collection of $N$ stabilizers, each of
which can have values $\pm 1$) into
$2^N$-component cavity cat-states, would seem to require an
unfortunate exponential increase in the required cat-size.
\end{widetext}

\end{document}